\documentclass[12pt,letterpaper]{article}
\pdfoutput=1
\usepackage{amsmath}
\usepackage{geometry}
\usepackage[none]{hyphenat}
\usepackage{fancyhdr}
\usepackage{enumitem}
\usepackage{graphicx}
\usepackage[labelfont=bf]{caption}
\usepackage{xcolor}
\usepackage{sidecap}
\usepackage{multicol}
\usepackage{wrapfig}
\usepackage{changepage}
\usepackage{cite}

\usepackage[labelfont={sf,bf}]{caption}
\begin{document}
\begin{figure*}[ht]
	\centering
	\includegraphics[width=\linewidth]{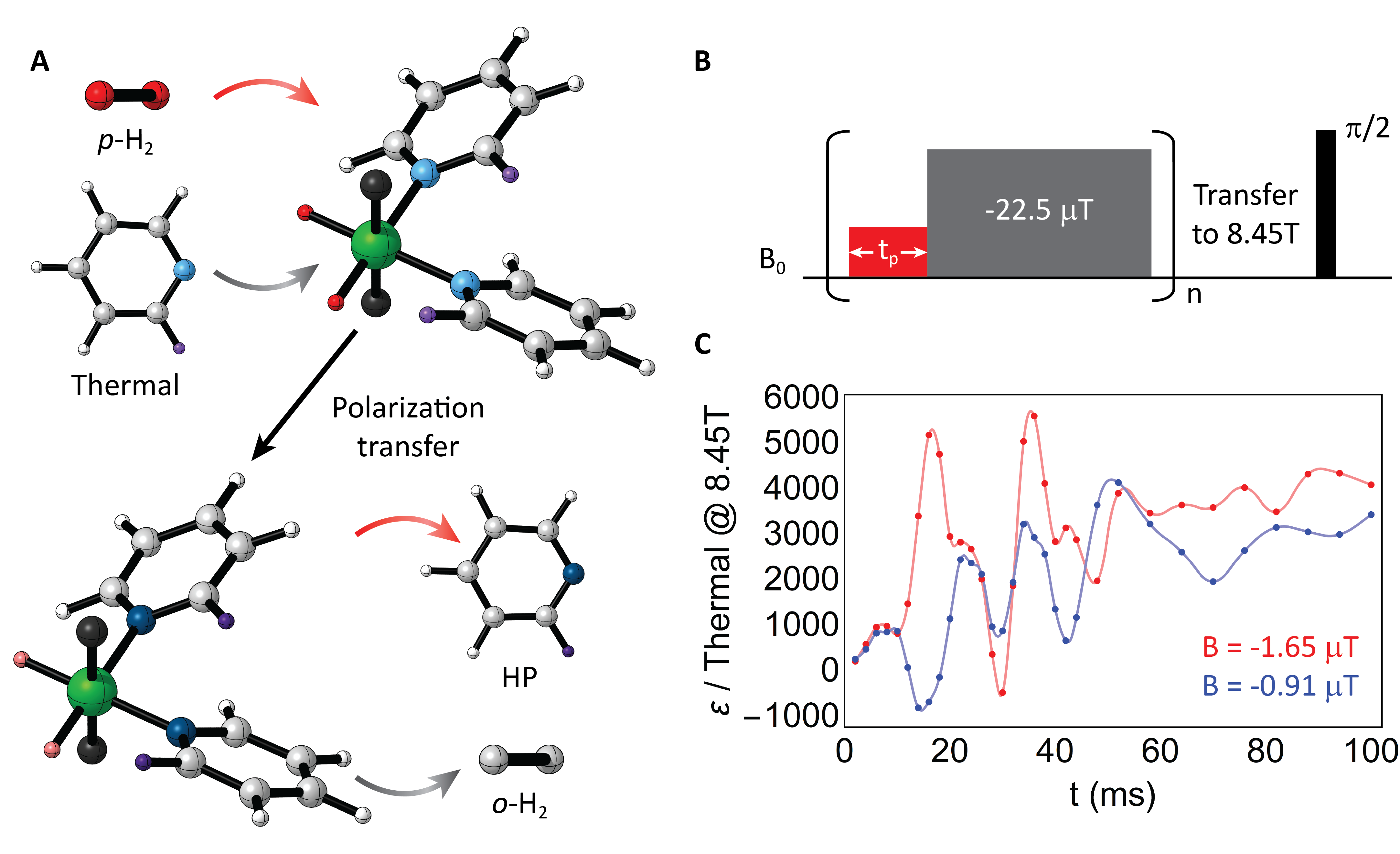}
	\caption{\textsf{Signal Amplification By Reversible Exchange, or SABRE, provides an ideal system to challenge the limits of an exchange model, given the complexity of the underlying dynamics. (\textbf{A}) SABRE transfers the singlet order of parahydrogen to a target ligand via reversible interactions with an iridium catalyst, and exhibits nonlinear dynamics that are highly dependent on the relative concentrations of each species. (\textbf{B}) The coherent hyperpolarization dynamics can be probed by interleaving pulses at or near the SABRE resonance condition (red) with periods far off resonance ($ B = -22.5 $ $\mu T $) to allow for exchange. (\textbf{C}) The ($ ^{15} $N,$ ^{13} $C)-acetonitrile SABRE system demonstrates rich dynamical information that varies with the resonant field as a result of a complex coupling network between $^{15} $N, $ ^{13} $C, and $^{1} $H. Lines are meant as a visual guide only.}} \label{fig:intro}
\end{figure*}

\sloppy
\begin{center}
{\noindent{\Large{\textbf{Infinite-order perturbative treatment for quantum evolution with exchange}}}}\\

\noindent Jacob R. Lindale$^1$, Shannon L. Eriksson$^{1,2}$, 
Christian P.N. Tanner$^1$, Warren S. Warren$^{3}$\\
{\textit{$^1$Department of Chemistry, Duke University, Durham, NC, 27708. 
$^2$School of Medicine, Duke University, NC, 27708. 
$^3$Department of Physics, Chemistry, Biomedical Engineering, and Radiology, Duke University, Durham, NC, 27708.}}
\end{center}
%
\begin{adjustwidth}{0.5in}{0.5in}
Many important applications in biochemistry, materials science, and catalysis sit squarely at the interface between quantum and statistical mechanics: coherent evolution is interrupted by discrete events, such as binding of a substrate or isomerization. Theoretical models for such dynamics usually truncate the incorporation of these events to the linear-response limit,  thus requiring small step sizes. Here, we completely re-assess the foundations of chemical exchange models and redesign a master equation treatment accurate to all orders in perturbation theory. The net result is an astonishingly simple correction to the traditional picture which vastly improves convergence with no increased computational cost. We demonstrate that this approach accurately and efficiently extracts physical parameters from highly complex experimental data, such as coherent hyperpolarization dynamics in magnetic resonance, and is applicable to a wide range of other systems.\\
\end{adjustwidth}
\begin{multicols}{2}
\noindent{\large{\textbf{Introduction}}}\\ 
Calculations of quantum evolution in dynamic systems, such as exchange or conversion between multiple discrete states, are important today in many disciplines \cite{RN2472,RN2481,RN2119,RN2138,RN2109}. Such calculations  first  became prominent in magnetic resonance more than fifty years ago with the McConnell equations \cite{RN2480}, which were introduced first as a classical approximation of the spin dynamics in exchanging systems. These equations could readily describe the dynamic spectra of uncoupled spin-1/2 nuclei, but were incapable of handling evolution under bilinear couplings. In contrast, the density matrix formalism\cite{RN2469,RN2482,RN2470} readily includes statistical averaging in the equilibrium state, and coherent evolution can be handled by unitary transformations involving calculation of a highly accessible propagator for spin systems.\\
\indent As it would be computationally impossible to explicitly calculate, for instance, the dynamics of $ 10^{20} $ nuclear spins, one averages over each molecule to form a reduced density matrix, wherein the form of ensemble interactions is obfuscated. Therefore, dynamic exchange effects require a more careful treatment of the expression of the ensemble action in the reduced density picture, modifying the time evolution from the form given by pure quantum mechanics ($ \partial_t \hat{\rho} = i\hbar^{-1}[\hat{\rho},\hat{\mathcal{H}}] $). The exchange interaction has been historically derived as an analog to the case of Redfield relaxation theory \cite{RN2469,RN556}, but the ensemble dynamics that generate relaxation occur on a timescale far faster than the evolution of the quantum degrees of freedom (fs-ps), effectively limiting the influence of these dynamics to the first observable moment. This is not valid for exchange, where it would be feasible for higher moments of the ensemble interaction to act on a timescale comparable to the coherent evolution.\\
%
\indent Despite the maturity of models for exchange, there is still considerable motivation to develop new methods to efficiently and accurately explore dynamic effects in systems undergoing quantum evolution. On the forefront of magnetic resonance techniques are hyperpolarization methods \cite{RN581,RN1851,RN573,RN996,RN505}, which overcome the intrinsically low signal-to-noise limits by distilling spin order from an external source. Of particular interest over the last decade is Signal Amplification By Reversible Exchange\cite{RN505,RN1564,RN1565,RN2272,RN1905,RN2483,RN2149,RN2477,RN1411,RN2462,RN1648,RN2109,RN2156,RN2131,RN1639,RN1518,RN2479,RN2063}, or SABRE, in which the singlet order of parahydrogen is converted into observable magnetization or more complex spin states on target ligands during transient interactions with an iridium catalyst (Figure \ref{fig:intro}A). Optimization of this technique requires accurate modeling of the system, which with the recent advent of coherently pumped SABRE experiments (Figure \ref{fig:intro}B) has revealed bizarre and complex dynamics\cite{RN2109,RN2474}. Accurately modeling the coherent hyperpolarization dynamics of systems like ($ ^{15} $N,$ ^{13} $C)-acetonitrile (Figure \ref{fig:intro}C) and subsequently fitting the experimental data has been impossible within previous frameworks for exchange, given the multitudinous exchange interactions, such as coligand exchange events and number of coupled spins (21 total spins).\\
\indent To that end, we completely re-interrogate the incorporation of dynamic exchange interactions in evolving quantum systems. We construct a reformulated dissipative master equation by recovering the traditional expression from the Dyson series and then continuing the derivation to infinite order in perturbation. The ramifications of extending the derivation of the dissipative master equation to all orders in the exchange interaction make a profound impact on the radius of convergence of exchange simulations with \textit{absolutely no additional computational cost} by deriving a simple scaling factor that accounts for all moments in the ensemble motion. In addition to the most general case of exchange between distinguishable ensembles, we show solutions for pseudorotation generated by Abelian groups of order 2 and 3 as well as for quantum dynamical selection, where coherent degrees of freedom alter the exchange interaction. By coupling this new infinite-order treatment exchange to a formulation of the exchange operators that scales linearly with the number of distinguishable ensembles, we can easily model highly complex systems that would be untenable within alternate exchange formalisms\cite{RN1648,RN2119}.\\

\noindent{\large{\textbf{Results}}}\\ 
Pursuing a traditional, master equation approach to chemical exchange requires the assumption that fundamentally discrete exchange events may be approximated as a continuous perturbation to the ensemble, shifting the model from assuming a Poisson process of a microcanonical ensemble to a Wiener process of a canonical ensemble. Quantum Monte Carlo (QMC) models discontinuously sample exchange events and are essentially exact, provided one can iterate the solution to convergence \cite{RN2109}. However, the cost of iterating a QMC solution to convergence would often make the calculation intractable on the timescale of experimental guidance. For instance, on the canonical 14-spin bis-($ ^{15} $N-pyridine) SABRE system, it would take approximately 300 years to run a single 60 second simulation at a modest exchange rate ($ k_{ex}=50 s^{-1} $) to approximately 99$ \% $ convergence.\\
\indent The methods to describe dynamic evolution in quantum systems are well established in the case of spin relaxation in liquids, which relies on the perturbative Dyson series expansion of the interaction-frame propagator. In this case, as well for the traditional case for exchange, the series is truncated to the leading term by assuming that the dynamic interaction is a small perturbation. The same result can be recovered by annealing the Liouville-von Neumann equation to the rate equations defining exchange by taking the tensor product between the quantum and chemical degrees of freedom. In both cases, this assumes that exchange acts linearly on the evolving quantum degrees of freedom, which is not well motivated. With our ansatz that discrete exchange events may be approximated as a continuous process, we recast the Dyson series for the case of exchange without any \textit{a priori} assumptions as to the magnitude or order of the exchange interaction.\\

\noindent{\textbf{\textit{Reformulation of the Dyson series for chemical exchange}}}\\
To begin, we partition the Hamiltonian into a stationary component $ \hat{\mathcal{H}}_0 $ and a stochastically modulated exchange interaction $ \hat{\mathcal{H}}_1(t) $, for which the equation of motion is given by $ (\hbar=1) $:
\begin{align}
\partial_t\hat{\rho}=-i\left[\hat{\mathcal{H}}_0+\hat{\mathcal{H}}_1(t),\hat{\rho}\right] \label{eq:1}
\end{align}
Boosting \eqref{eq:1} into the interaction frame gives
\begin{align}
\partial_t\hat{\sigma}=-i\left[\hat{\tilde{\mathcal{H}}}_1(t),\hat{\sigma}\right]\!, \label{eq:2}
\end{align}
where we employ the convention $ \hat{\rho}\rightarrow\hat{\sigma} $ for distinguishability between the representations and $ \hat{\tilde{\mathcal{H}}}_1(t) \equiv \exp(-i \hat{\mathcal{H}_0}t)\hat{\mathcal{H}}_1(t)\exp(i \hat{\mathcal{H}_0}t) $ is the interaction representation denoted by the tilde. Equation \eqref{eq:2} may then be formally integrated and iteratively substituted back into the expression to generate the Dyson series. When doing so, we assume that the correlation time of the exchange interaction is much faster than the coherent evolution of the system, which allows us to write $ \hat{\sigma}(t')=\hat{\sigma}(t) \forall t' $ and extend the upper limit of integration to infinity:
\begin{align}
\nonumber \partial_t\hat{\sigma}(t)&=-i\left[\hat{\tilde{\mathcal{H}}}_1(t),\hat{\sigma}(t)\right]\\
&-\vec{\mathcal{T}} \int_{0}^{\infty}dt' \left[\hat{\tilde{\mathcal{H}}}_1(t),
\left[\hat{\tilde{\mathcal{H}}}_1(t'),\hat{\sigma}(t)\right]\right]+\cdots \label{eq:3}
\end{align} 
In this equation, $ \vec{\mathcal{T}} $ is the Dyson time ordering operator, which imposes $ t > t'$. At this juncture, we introduce $ \hat{\tilde{\mathcal{H}}}_1(t) $ as the operator expansion
\begin{align}
\hat{\tilde{\mathcal{H}}}_1(t)\equiv\sum_{q}\hat{F}_q(t)\hat{\tilde{A}}_q, \label{eq:4}
\end{align}
where $ q $ indexes through uncoupled exchange mechanisms, $ \hat{F}_q(t) $ are real, stochastic operators describing the time evolution of exchange, and $ \hat{\tilde{A}}_q $ define the interaction of exchange with the evolving quantum system. We assume the system is in a chemical steady state, and therefore $ \langle\hat{F}_q(t)\rangle=0 $ and is importantly not the first moment of the exchange rate. Furthermore, this has the repercussion that all odd-order terms in the expansion necessarily average to zero, ensuring that chemical exchange generates no complex phase rotations in $ \hat{\sigma} $. Substituting equation \eqref{eq:4} into \eqref{eq:3} and ensemble averaging gives the leading term:
\begin{align}
\nonumber \partial_t\hat{\sigma}=
-\vec{\mathcal{T}}\sum_{pq}\int_{0}^{\infty}dt'\! &\hat{\hat{A}}_pe^{-i\hat{\mathcal{H}}_0(t'-t)}\hat{\hat{A}}_qe^{i\hat{\mathcal{H}}_0(t'-t)}\hat{\sigma}\\
&\times \langle\hat{F}_p(t)\hat{F}_q(t')\rangle \label{eq:5}
\end{align}
Note that we have shifted to the commutation superoperator representation $ ([\hat{O},\bullet]\equiv\hat{\hat{O}}) $ and have dropped the formal time-dependence of $ \hat{\sigma} $ as well as the tilde-notation for the interaction frame for legibility. The integrand is then the correlation function of the ensemble motion, which for a stochastic, real-valued, and time-continuous (Wiener) process is delta-correlated in time. To avoid violating the time-ordering operator, $ \vec{\mathcal{T}} $, only self-correlated terms can give non-zero amplitudes upon integration, which may be accounted for by imposing $ \delta_{pq} $. Importantly, the delta-correlation imposes that $ \exp(i\hat{\mathcal{H}}_0(t'-t)) $ is unity, making all interaction-frame superoperators for exchange identical to their lab space representation. Then, the time-ordered integral simply determines the rate of the exchange process, which is the probability of exchange given a characteristic lifetime $ (\tau_q) $ during a finite period of time, $(\Delta t \tau_q^{-1})/\Delta t $. Upon integration, this gives:
\begin{align}
\nonumber \vec{\mathcal{T}} \int_{0}^{\infty}dt' \langle\hat{F}_p(t)\hat{F}_q(t')\rangle&=
\int_{0}^{\infty}dt'\delta(t-t')\delta_{pq}\frac{\Delta t \tau_q^{-1} }{\Delta t}\\
&=\frac{1}{\tau_q}
 \label{eq:6} 
\end{align}
The term $\hat{\hat{A}}_p\hat{\hat{A}}_q $ can now be written as $\hat{\hat{A}}_q^2$, which is immediately realized in its more common notation as the exchange superoperator $-\hat{K}_q $, where the sign convention arises to ensure that exchange forms a completely positive map. Together with equation \eqref{eq:6}, this returns the canonical form of the chemical exchange master equation, which in the Schrödinger representation is:
\begin{align}
\partial_t\hat{\rho}=-i\left[\hat{\mathcal{H}}_0,\hat{\rho}\right]
+\sum_{q} \frac{\hat{K}_q\hat{\rho}}{\tau_q} \label{DME2}
\end{align}
\indent Truncating the expansion here essentially imposes that exchange only acts linearly on $ \hat{\rho} $ during the finite period of time $ \Delta t $ (ie. the simulation step size), as this was the condition under which equation \eqref{eq:6} was defined. In only the simplest cases can equation \eqref{DME2} be homogenized and analytically integrated, wherein the finite period of time $ \Delta t \rightarrow dt$ and the assumption that an exchange interaction of any magnitude acts linearly on the quantum degrees of freedom is accurate. However, now that we have explicitly established the derivation of the chemical exchange master equation from the Dyson series, it is simple to continue the derivation to higher order terms. Remembering that all odd-order terms necessarily go to zero upon ensemble averaging, the next non-zero interaction in the expansion arrives in the fourth-order term, which after substitution of \eqref{eq:4} and realizing that there can be only powers of $ \hat{\hat{A}}_q^2 $ to give rise to the superoperator $ \hat{K}_q $ is:
\begin{align}
\nonumber &\partial_t\hat{\sigma}=
\sum_{pq} \hat{\hat{A}}_p^2\hat{\hat{A}}_q^2\hat{\sigma}\\
&\times \vec{\mathcal{T}} \int_{0}^{\infty}dt'\int_{0}^{t'}dt''\int_{0}^{t''}dt''' \langle\hat{F}_p(t)\hat{F}_p(t')\hat{F}_q(t'')\hat{F}_q(t''')\rangle \label{eq:8}
\end{align}
The integrand of equation \eqref{eq:8} is a four-point correlator that may be factored into a sum of two-point correlator products by Isserlis' theorem, where the form of each correlator is given by equation \eqref{eq:6}. There are $ (n-1)!! $ identical terms for an \textit{n}-point correlator after factorization when the process is $ \delta $-correlated, where $ n!! $ is the semifactorial of $ n $, defined as the factorial using only integers of the same parity as $ n $ $ (5!!=5\times3\times1) $. Additionally, given the time-symmetry of the Wiener process, there will be $ (n-1)! $ degenerate time orderings upon integration, accounted for with division by of the correlator amplitude by the degeneracy. Integration then gives:
\begin{align}
\nonumber \vec{\mathcal{T}} \int_{0}^{\infty}dt'&\int_{0}^{t'}dt''\int_{0}^{t''}dt''' \langle\hat{F}_p(t)\hat{F}_p(t')\hat{F}_q(t'')\hat{F}_q(t''')\rangle\\
&=\frac{(4-1)!!}{(4-1)!}\left(\frac{\Delta t}{\tau_q}\right)^2/\Delta t
=\frac{\Delta t}{2\tau_q^2} \label{eq:9}
\end{align}
Notice that $ \delta_{pq} $ prevents any cross terms between $ p $ and $ q $ from arising in the summation, hence why integration generates a rate proportional to $ \tau_q^2 $. Equation \eqref{eq:8} then becomes:
\begin{align}
\partial_t\hat{\sigma}=
\sum_{q} \left(-\hat{\hat{A}}_q^2\right)^2
\frac{\Delta t}{2\tau_q^2}\hat{\sigma} \label{eq:10}
\end{align}
The fourth-order term describes the probability of two exchange events occurring during a finite period of time, as the exchange interaction can be expanded into successive applications of $ -\hat{\hat{A}}_q^2 $. We shall define the powers of the exchange interaction conditioned to specific cases and otherwise leave it in its more general form. Using the assumptions established here, it is then beneficial to rewrite the entire Dyson series for exchange as:
\begin{align}
\partial_t\hat{\sigma}=
\sum_{q}\left\lbrace\frac{1}{\tau_q}\sum_{k=0}^\infty \left(-\hat{\hat{A}}_q^2\right)^{k+1}
\frac{1}{k!}\left(\frac{\Delta t}{2\tau_q}\right)^k\right\rbrace\hat{\sigma} \label{DMEgen}
\end{align}
Equation \eqref{DMEgen} may be simplified by establishing a more rigorous definition of the exchange operator $ -\hat{\hat{A}}_q^2 $, which will show for a general case as well as more specific applications. Before doing so, it is pertinent to note that the general operator action of \eqref{DMEgen} can be written
\begin{align}
\nonumber \left(-\hat{\hat{A}}_q^2\right)^{k+1}\hat{\sigma}&=\left(-\hat{\hat{A}}_q^2\right)^{k}\hat{K}_q\hat{\sigma}\\
&=\gamma \hat{K}_q\hat{\sigma}
\label{eq:12}
\end{align}
where the first equality is inherent given the definition of $ \hat{K}_q $ and the second is possible if $ \hat{K}_q\hat{\sigma} $ are eigenfunctions of $ (-\hat{\hat{A}}_q^2)^k $, where $ \gamma $ is then a constant. Under this condition, the infinite sum in equation \eqref{DMEgen} would be independent of the interaction superoperator and evaluation of all moments of the exchange interaction would have no additional computational cost over a traditional formulation. \\

\noindent{\textit{Exchange between distinguishable ensembles}}\\
The most general formulation for exchange is to form a composite vector space constructed from the direct sum of $ m $ discrete chemical configurations that form manifolds of quantum states, where exchange allows flow of populations between manifolds via projection operations. This formulation is unrestricted in the systems that it may describe, as it is trivial to project between systems of different sizes and projections may both encapsulate linear and non-linear contributions to the evolution. When constructed carefully, this method grows linearly in cost with $ m $, as projections need not act on the entire composite vector space.\\
\indent We find the form of the exchange interaction for this case by recognizing that projections are idempotent operations ($ \gamma =1 $ in equation \eqref{eq:12}), such that:
\begin{align}
\nonumber \left(-\hat{\hat{A}}_q^2\right)^{k+1}&=\left(-1\right)^{k+1}\left(\hat{\hat{A}}_q^2\right)^{k+1}\\
\nonumber 
&=\left(-1\right)^{k+1}\left(-\hat{K}_q\right)^{k+1}\\
&=\left(-1\right)^{k}\hat{K}_q
\label{eq:13}
\end{align}
Note that we have re-indexed the alternating term for convenience. There is an intricate ramification of equation \eqref{eq:13}, in that the group $ \mathcal{G}_q $ containing all powers of the exchange superoperators for a given system is isomorphic to its coset of linear-power superoperators $ \mathcal{S}_q^{(1)} $, which equivalently form the kinetic equations for the chemical dynamics. As such, considering all moments in the exchange interaction will only ever generate dynamics that are directly reflected in $ \hat{K}_q $. Given \eqref{eq:13} and that the summation over $ k $ is simply the Maclaurin series for the exponential, equation \eqref{DMEgen} may be written in the Hilbert space as:
\begin{align}
\partial_t\hat{\rho}=-i\left[\hat{\mathcal{H}}_0,\hat{\rho}\right]
+\sum_{q}\left\lbrace\frac{\hat{K}_q}{\tau_q}\,\textcolor{red}{\exp\left(\frac{-\Delta t}{2 \tau_q} \right)}\!\right\rbrace\hat{\rho}
 \label{DMEx1}
\end{align}
In this form, equation \eqref{DMEx1} is the \textit{exact, closed form} solution of quantum evolution with exchange, which we call the Exact Dissipative Master Equation, or DMEx. 
In its exact form, the only difference that arises in comparison to the traditional form is the exponential factor highlighted in red, where it is clear that in the limit where $ \Delta t \rightarrow dt $, the equation converges back to \eqref{DME2}. That arises as the impact of higher moments in exchange go away over small periods of time, because it is impossible for multiple exchange events to occur simultaneously in the limit of infinitesimal step sizes. However, as $ \Delta t $ becomes larger, the higher order terms account for moments in the dynamics of the ensemble that are not present when one assumes a linear coupling between the quantum and exchange degrees of freedom.\\

\noindent{\textit{Pseudorotation by Abelian permutation groups}}\\
While equation \eqref{DMEx1} is valid for any exchanging quantum system, it would be inconvenient to expand a system into separate manifolds when the coherent evolution within those manifolds, or a subset of those manifolds, is identical. This is the case for exchange generated by pseudorotation, such as the canonical example of cyclohexane inversion in magnetic resonance. As such, it is convenient to recast equation \eqref{DMEgen} for the cases of pseudorotation generated by 2- and 3-fold symmetric permutation groups, which we shall call $ \mathcal{G}^2 $ and $ \mathcal{G}^3 $ pseudorotations, respectively. As we are recasting the DMEx for specific cases, we will drop the $ q $-index and write the explicit form of the equation of motion. It is important to note that all of the assumptions made to derive equation \eqref{DMEgen} remain valid for these conditions.\\
\indent In the case of pseudorotations, which contain inherently coupled exchange processes, we define the first moment of the exchange interaction operator as:
\begin{align}
\nonumber -\hat{\hat{A}}^2\equiv\hat{\tilde{K}}&=\frac{1}{2}\left(\hat{R}\otimes\hat{R}^{-1}+
\hat{R}^{-1}\otimes\hat{R}\right)-\hat{E}\\
&=\frac{1}{2}\left(\hat{K}+\hat{K}^\dagger\right)
\label{eq:15}
\end{align}
The operators $ \hat{R}^{\pm1}\otimes\hat{R}^{\mp1} $ generate the forward $ (\hat{K}) $ and backwards  $ (\hat{K}^\dagger) $ rotations, which are coupled with equivalent probability. We have also written the exchange superoperator $ \hat{K} $ in a traceless form for convenience instead of writing a separate term proportional to unity ($ \hat{E} $), as done in the general case. The forward and backwards rotations are equivalent for $ \mathcal{G}^2 $ pseudorotations, which allows \eqref{eq:15} to be reduced to:
\begin{align}
\hat{\tilde{K}}=\hat{R}\otimes\hat{R}^{-1}-\hat{E} \label{eq:16}
\end{align}
The form of the higher powers of the exchange interaction are given as
\begin{align}
\left(-\hat{\hat{A}}^2\right)^{k+1}=(-2)^{k}\hat{\tilde{K}}, \label{eq:17}
\end{align}
and hence give the DMEx for $ \mathcal{G}^2 $ pseudorotation in the Hilbert space:
\begin{align}
\partial_t\hat{\rho}=-i\left[\hat{\mathcal{H}}_0,\hat{\rho}\right]
+\frac{\hat{R}\hat{\rho}\hat{R}^{-1}-\hat{\rho}}{\tau}\,\textcolor{red}{\exp\left(\frac{-\Delta t}{\tau} \right)}
\label{DMExG2}
\end{align}
Again the term highlighted in red returns the canonical equation of motion for exchange in magnetic resonance when taken to the limit of analytical integration ($ \Delta t \rightarrow dt $). Note that in this case, the argument of the exponent is proportional to $ 1/\tau $, as opposed to the other cases presented here. This arises as a ramification of the definition that was employed for $ \hat{\tilde{K}} $.\\
\indent The case of $ \mathcal{G}^3 $ pseudorotations no longer permits the reduction of equation \eqref{eq:15} to \eqref{eq:16}, as $ \hat{K}\ne\hat{K}^\dagger $. It is pertinent to note the relation $ \hat{R}^2\otimes\hat{R}^{-2}=\hat{R}^{-1}\otimes\hat{R} $, which allows for any quadratic or higher rotation to be written in terms of the linear term. The higher powers of the exchange interaction then reduce to the simple form:
\begin{align}
\left(-\hat{\hat{A}}^2\right)^{k+1}=(-1)^{k}\hat{\tilde{K}} \label{eq:19}
\end{align}
Plugging equation \eqref{eq:19} into \eqref{DMEgen} gives the DMEx for $ \mathcal{G}^3 $ pseudorotation as
\begin{align}
\partial_t\hat{\rho}=-i\left[\hat{\mathcal{H}}_0,\hat{\rho}\right]
+
\frac{\hat{\tilde{K}}\hat{\rho}}{\tau}\,\textcolor{red}{\exp\left(\frac{-\Delta t}{2\tau} \right)},
\label{DMExG3}
\end{align}
where
\begin{align}
\hat{\tilde{K}}\hat{\rho}=\frac{1}{2}\left(\hat{R}\rho\hat{R}^{-1}
+\hat{R}^{-1}\rho\hat{R}\right)-\hat{\rho}.
\label{eq:21}
\end{align}
Again, the only difference between the exact treatment and the traditional implementation for exchange is the exponential term. It is important to note that equations \eqref{DMExG2} and \eqref{DMExG3} can be used alone or in conjunction with the more general DMEx formalism shown in equation \eqref{DMEx1} as a method of contracting the composite vector space, with an obvious example being to use equation \eqref{DMExG3} to compress the manifolds corresponding to the three orientations of methyl rotation with magnetically inequivalent interactions. This does require that the rate connecting the contracted manifolds to any other manifold be identical, otherwise the contraction is invalid.\\

\noindent{\textit{Quantum dynamical selection}}\\ 
An interesting case to examine is when the quantum and chemical dynamics are coupled, such as in quantum dynamical selection (QDS), where quantum evolution dictates the evolution of the exchange degrees of freedom. In this case, we must re-assess the assumptions made to arrive at equation \eqref{DMEgen} pertaining to the stochastic motion of the ensemble. However, we will construct this case as exchange between distinguishable ensembles, leaving equation \eqref{eq:13} intact. Importantly, this case is restricted by the condition that
\begin{align}
\left[\hat{F}_q(t),\hat{\sigma}\right]\ne 0, \label{eq:22}
\end{align}
such that we may no longer ensemble average $ \hat{\sigma} $ and the ensemble motion operators $ \hat{F}_q $ separately. This gives rise to the leading term in the Dyson series:
\begin{align}
\partial_t\hat{\sigma}=
-\sum_{pq} \hat{\hat{A}}_p\hat{\hat{A}}_q
\vec{\mathcal{T}} \int_{0}^{\infty}dt' \langle\hat{F}_p(t)\hat{F}_q(t')\hat{\sigma}\rangle \label{eq:23}
\end{align}
The factorization of the three point correlator generates
\begin{align}
\nonumber &\langle\hat{F}_p(t)\hat{F}_q(t')\hat{\sigma}\rangle=
\langle\hat{F}_p(t)\hat{F}_q(t')\rangle \langle\hat{\sigma}\rangle\\&+
\langle\hat{F}_p(t)\hat{\sigma}\rangle \langle\hat{F}_q(t')\rangle+
\langle\hat{F}_q(t')\hat{\sigma}\rangle \langle\hat{F}_p(t)\rangle \label{eq:24}
\end{align}
As $ \langle\hat{F}_q\rangle=0 $ is retained with the assumption of a chemical steady state, only the first term in the factorization is retained. Furthermore, as the ensemble is still macroscopically described by a Wiener process, the $ \delta $-correlation is retained and $ \delta_{pq} $ is imposed to avoid violating the time-ordering. For any $ (2n+1) $-point correlator, only the single term that averages $ \hat{\sigma} $ separately will be retained. If we then insert equation \eqref{eq:24} into \eqref{eq:23} and define
\begin{align}
\vec{\mathcal{T}} \int_{0}^{\infty}dt' \langle\hat{F}_q(t)\hat{F}_q(t')\rangle \equiv \hat{\hat{\Phi}}_{q}, \label{eq:25}
\end{align} 
where $ \hat{\hat{\Phi}}_{q} $ is a ensemble motion superoperator that determines the rate of the process as a function of the population in the quantum state coupled to the exchange process, we obtain as the equation of motion:
\begin{align}
\partial_t\hat{\sigma}=
-\sum_{q} \hat{\hat{A}}_q^2\hat{\hat{\Phi}}_{q}\hat{\sigma} \label{eq:26}
\end{align}
We may impose that
\begin{align}
\hat{\hat{\Phi}}_{q}\hat{\sigma}=\zeta_q\hat{\sigma} \label{eq:27}
\end{align} 
such that $ \zeta_q $ is a constant that is the rate of the ensemble motion dictated by $ \hat{\hat{A}}_q$, which states that $ \hat{\sigma} $ is an eigenstate of $ \hat{\hat{\Phi}}_{q} $ at all times. This relation is validated when $ \delta_{pq} $ is imposed, which prevents the $ \hat{\hat{\Phi}}_{q}$ from generating a tensor of rank 1 or higher. As such, we may write the form of equation \eqref{eq:25} as
\begin{align}
\nonumber \hat{\hat{\Phi}}_{q}&=\vec{\mathcal{T}} \int_{0}^{\infty}dt' \frac{\delta(t-t')}{\Delta t/(\Delta t \tau_q^{-1})}\mathcal{P}_q\\
&=\frac{\mathcal{P}_q}{\tau_q}, \label{eq:28}
\end{align} 
where $ \mathcal{P}_q $ scales the rate by the projection of the quantum state which couples to the exchange process and is a constant. As projection operators are idempotent, we may derive the DMEx for QDS directly from the derivation of the general case, with an additional factor of $ \mathcal{P}_q $ that makes the rates time-dependent; in the Hilbert space, this is:
\begin{align}
\partial_t\hat{\rho}=-i\left[\hat{\mathcal{H}}_0,\hat{\rho}\right]
+\sum_{q}\left\lbrace\frac{\hat{K}_q\mathcal{P}_q}{\tau_q}\,\textcolor{red}{\exp\left(\frac{-\Delta t \mathcal{P}_q}{2 \tau_q} \right)}\right\rbrace\hat{\rho}
\label{DMExQDS}
\end{align}
Expressing the exchange rates as time-dependent quantities complicates interpretation of this case. However, if we choose to express the ensembles using both chemical and quantum degrees of freedom, we recover equation \eqref{DMEx1} as the equation of motion for exchange. The exchange rates between the redefined manifolds are then time-independent.\\

\begin{figure*}[ht]
	\centering
	\includegraphics[width=\linewidth]{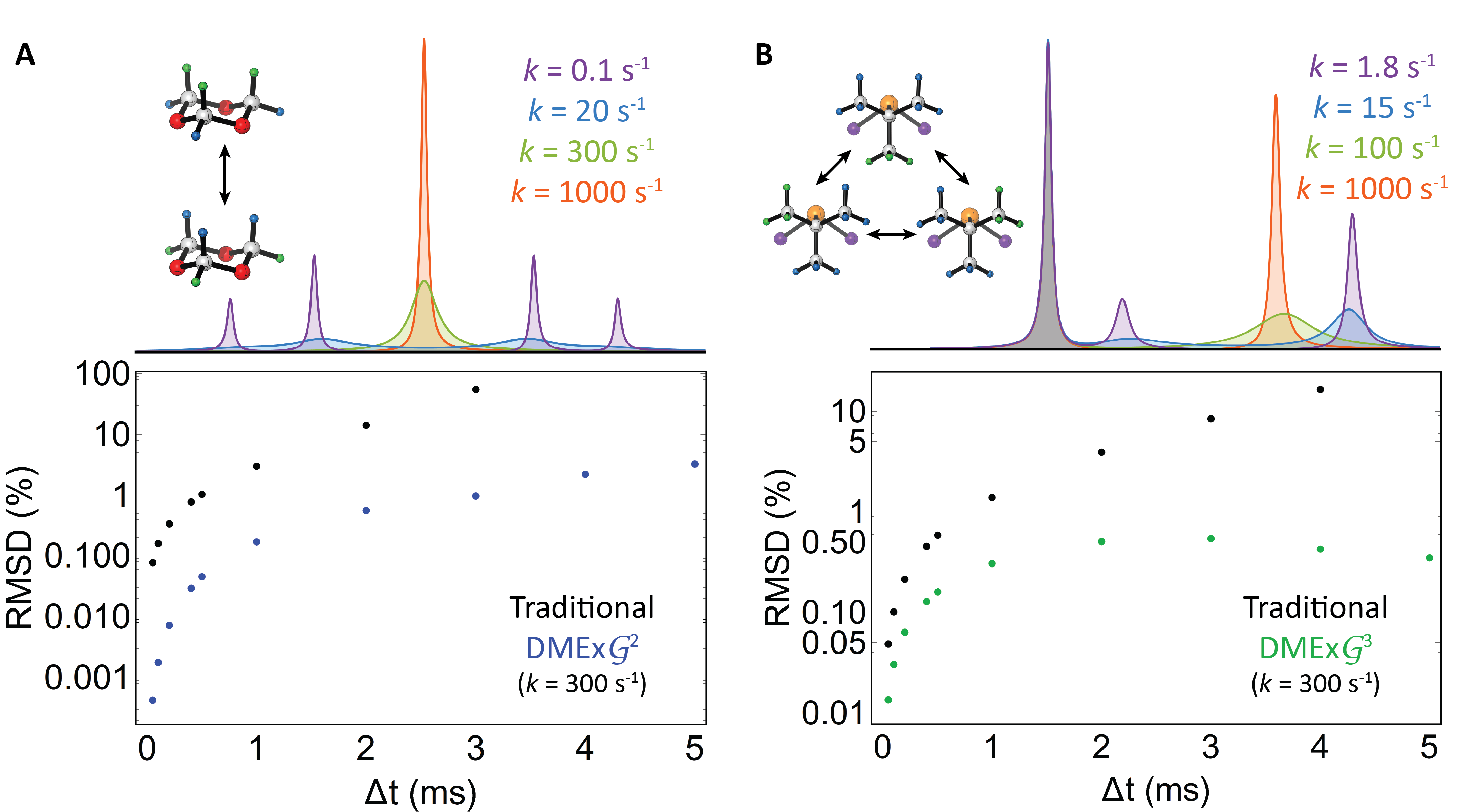}
	\caption{\textsf{Example simulations and convergence of DMEx models for $ \mathcal{G}^2 $ and $ \mathcal{G}^3 $ pseudorotation. The model systems are s-trioxane ring inversion (\textbf{A}) and tert-butyl rotation in t-BuPCl$_2 $ (\textbf{B}). Spectra were calculating using $ \Delta t=1 $ ms.}} \label{fig:error G2/G3}
\end{figure*}

\begin{figure*}[!htb]
	\centering
	\includegraphics[width=0.97\linewidth]{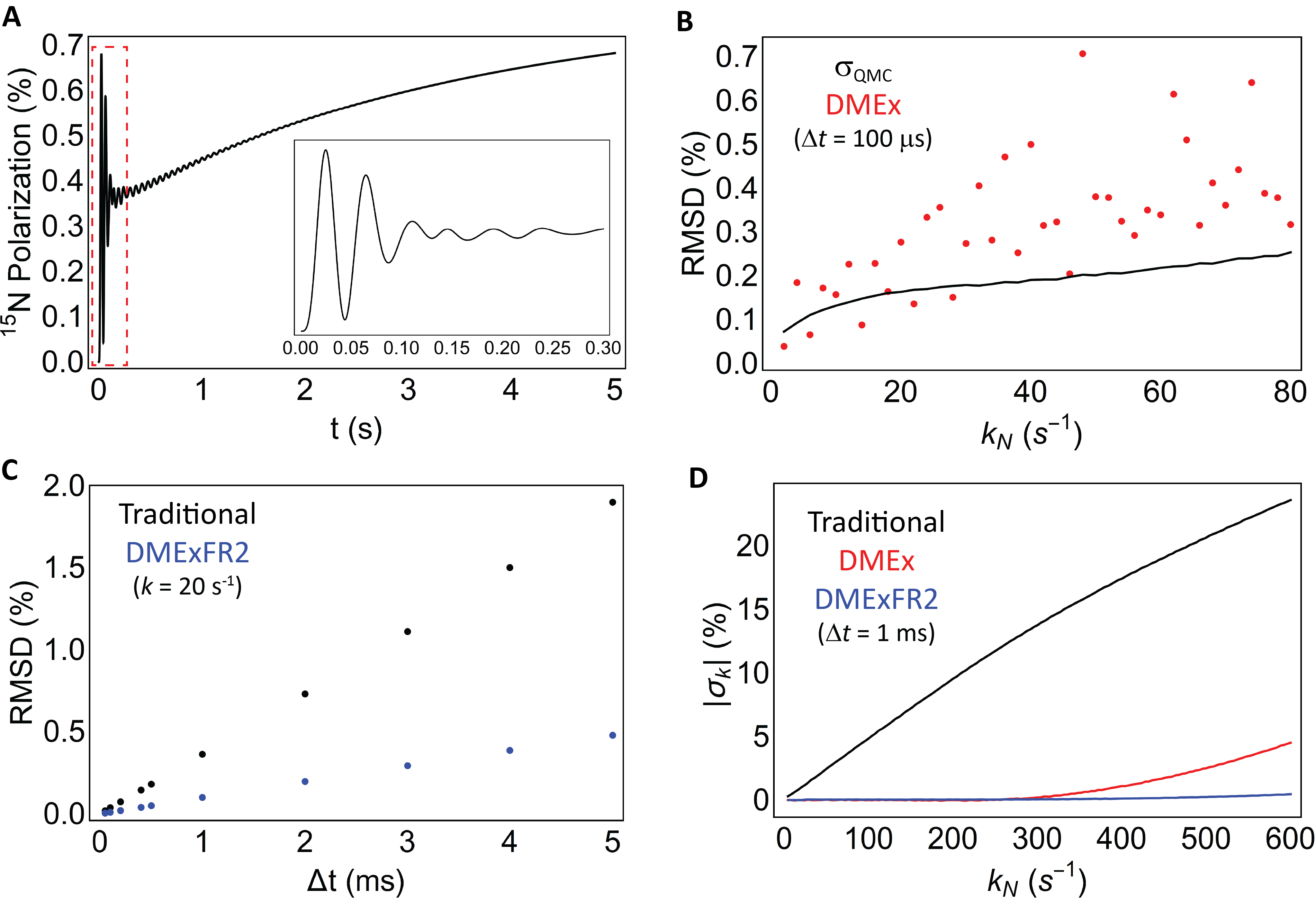}
	\caption{\textsf{Example DMEx simulations of complex systems using $ ^{15} $N SABRE as a model. (\textbf{A}) An example of SABRE hyperpolarization dynamics is shown for reference, calculated on a 6-spin $ ^{15} $N SABRE-SHEATH system. (\textbf{B}) There is a significant error in the DMEx when compared to a QMC solution at the same exchange rate (red), using the convergence error in the QMC solution as a threshold (black) with an average of $ (0.142\pm0.018)\% $. (\textbf{C}) Even with nonlinear effects incorporated in the simulation, the DMExFR2 exhibits a larger radius of convergence over the traditional implementation by approximately a factor of 4. (\textbf{D}) Scanning $ \sigma_k $ as a function of the exchange rate highlights that there is no stability in the self-consistency of the traditional implementation, whereas for the DMEx and DMExFR2, the solutions are self consistent even when the exchange rate approaches the step size.}} \label{fig:error SABRE}
\end{figure*}

\noindent{\textbf{\textit{Performance of DMEx models}}}\\
The chemical exchange master equation is only homogeneous and analytically integrable in the simplest of cases and can acquire nonlinearities when one considers reversible exchange between distinguishable ensembles. Therefore, we will evaluate the performance of DMEx methods using sympletic ``leapfrog" integration as an approximate solution to the equations of motion, which in the simplest case looks like:
\begin{align}
\nonumber &\hat{\rho}(t+\Delta t)=\hat{\mathcal{U}}^\dagger\hat{\rho}(t)\hat{\mathcal{U}}\\
&+\sum_q\frac{\Delta t}{\tau_q} \exp\left(\frac{-\Delta t}{2\tau_q}\right)
\left(\hat{K}_q-\hat{E}\right)\left(\hat{\mathcal{U}}^\dagger\hat{\rho}(t)\hat{\mathcal{U}}\right) \label{eq:DMEx solution}
\end{align}
In this equation, $ \hat{\mathcal{U}} \equiv \exp(i\hat{\mathcal{H}}_0\Delta t) $. This is an ideal computational method as it only involves forward propagation of the solution, requires the fewest number of matrix operations, and produces linear evolution under the spin Hamiltonian and evolution to all orders in the exchange interaction. This method has a small intrinsic error associated with the solution, in that the first step only evolves quantum mechanically. A more accurate way to solve the equation of motion would be to evolve the initial density matrix backwards in time by $ \Delta t/2 $ and then utilizing equation \eqref{eq:DMEx solution} to generate the solution. Doing so shifts the actions of  $ \hat{\mathcal{U}} $ and $ \hat{K}_q $ by a half-step and corrects for this initial error. However, we have found that this makes little difference in the solution, thus we retain equation \eqref{eq:DMEx solution} so to avoid generating a non-integral number of steps. To isolate errors arising from exchange, we have constructed all of the following simulations in the Hilbert space of the system, where one can exactly evaluate quantum evolution in systems up to 15 coupled spins.\\
\indent Dynamic NMR spectra under pseudorotation have been studied and understood for decades . Spectral features are well resolved in the limit of slow exchange, which broaden and coalesce as the exchange rate increases, and ultimately result in line narrowing in the fast exchange limit. This is reflected in spectra of s-trioxane\cite{RN2470} undergoing ring-inversion (Fig. \ref{fig:error G2/G3}A), where the axial (blue) and equatorial (red) have different chemical shifts and the geminal $ ^2J_{HH} $ coupling is observable. As exchange increases, the spectrum collapses to a singlet, as the axial and equatorial positions become, on average, magnetically equivalent. Similar effects appear for the tert-butyl rotation in t-BuPCl$_2 $  (Fig. \ref{fig:error G2/G3}B), which additionally exhibits a transition that is invariant under exchange and thus does not broaden \cite{RN2484}.\\
\indent For either of these systems, the pseudorotation matrices are generated by expressing a spin label permutation matrix in the appropriate basis. For convenience we will use the Zeeman basis in this example. In the case of s-trioxane, where the axial and equatorial protons interchange, it is most convenient to setup the system such that axial protons have odd indices and equatorial protons have even indices. Then, the rotation $ \hat{R} $ is given by:
\begin{align}
\hat{R}&=\hat{P}_{12}\otimes\hat{P}_{34}\otimes\hat{P}_{56}\\
\hat{P}_{ij}&=
\begin{bmatrix}
1&0&0&0\\
0&0&1&0\\
0&1&0&0\\
0&0&0&1\\
\end{bmatrix}
\end{align} 
$ \hat{P}_{ij} $ is the permutation matrix that interchanges the $ |\alpha \beta\rangle $ and $ |\beta\alpha\rangle $ states, whereas the $ |\alpha \alpha\rangle $ and $ |\beta\beta\rangle $ states are invariant under exchange. Using this method, it is trivial to arbitrarily re-index the entire system and is computationally efficient, because the transformation from the original basis to the re-indexed basis is unitary.\\
\indent When calculating these spectra, we find that the traditional implementation and the DMEx converge to the same solution as $ \Delta t \rightarrow dt $. However, the DMEx exhibits a significantly smaller error at any step size than the traditional implementation, and only accrues an error on the order of $ 1\% $ when the step size \textit{exceeds} the average lifetime. In this limit, the traditional implementation loses stability and the trace of $ \hat{\rho} $ deviates from unity.This immediately provides the ability to take larger step sizes with the DMEx implementation. In the case of s-trioxane, an error in the solution of $ \approx 1\% $ requires $ \Delta t=1  $ ms in the DMEx and $ \Delta t=0.1$ ms using the traditional implementation, thus requiring one to sample far fewer data points. In considering all moments of the exchange interaction, the radius of convergence of the Dyson expansion is far larger than it would be by assuming conditions similar to those used for exchange.\\
\indent While these model systems provide illustrative examples of the performance of the DMEx model, they are far from the more challenging cases in dynamic systems. As noted previously, an interesting system that has gained much attention in the past decade is the hyperpolarization method SABRE, wherein large non-thermal nuclear magnetization is distilled from parahydrogen via reversible interactions with an iridium catalyst. Current efforts are focused on optimizing the extraction of spin order from parahydrogen, which requires accurate modeling of the quantum and exchange dynamics in realistic systems. For reference, an example simulation of the coupled coherent and exchange dynamics that drive SABRE hyperpolarization is shown in Figure \ref{fig:error SABRE}A, where the evolution of the $ ^{15} $N polarization is calculated under the experimental conditions for SABRE-SHEATH.\\
\indent In deriving the DMEx, we began with the ansatz that exchange could be considered as a time-continuous perturbation of the ensemble, but it is interesting to see when this assumption fails. The perturbation generated by exchange in the slow exchange limit is small, allowing the solution to be largely dictated by the quantum dynamics, and conversely in the fast exchange limit, quantum evolution cannot generate large excursions from equilibrium when constantly disrupted by exchange. In the intermediate regime where SABRE exists, characterized by exchange rates on the order of the dominant couplings, it is no longer trivial to motivate that large excursions from equilibrium would not be impactful on the dynamics. To probe this, we compared our previous Quantum Monte Carlo (QMC) model for SABRE against the DMEx on a 3-spin SABRE system (Fig. \ref{fig:error SABRE}B) with a dominant coupling of $ ^2J_{NH}= -24 $ Hz. In this regime, there is a significant difference between the convergence error of the QMC solution ($ \sigma_{QMC} $) and the DMEx solution, however this error is on average only $ (0.142\pm0.018)\% $. It is important to note that this analysis is limited to the smallest systems given the large cost of iterating the QMC solution, and the error accrued by the DMEx is negligible on simulation timescale relevant to experimental guidance.\\
\indent When modeling more complex systems, such as those often found in SABRE, it is critical for the cost of the DMEx to be augmented with an efficient method for exploring complex interactions, otherwise circumventing the benefits of an infinite-order treatment by excessive computational costs.  In SABRE, these interactions include quantum evolution of multiple species, rebinding of previously polarized ligands to the activated complex, binding site competition with spin-inert coligands, and relaxation. We call this SABRE-specific model the ``DMExFR2'' to indicate free ligand, rebinding, and relaxation effects are included. The most efficient way of accomplishing an efficient implementation of exchange is by expressing the interactions as block diagonal with respect to individual manifolds, which we call ``manifold-diagonal'' for simplicity and will motivate using the example of SABRE.\\
\indent In SABRE, we primarily consider two different species: one in which the hyperpolarization target is bound to the iridium, which we call ``bound species'', and another in solution, which we call ``free species''. Coherent evolution in the manifolds is established by separately propagating a bound species density matrix ($\hat{\rho}_{bS} $) and the dissociated free species density matrix ($\hat{\rho}_{fS} $) under their respective nuclear spin Hamiltonians:
\begin{align}
\nonumber \partial_t\hat{\rho}_{fS}&=
-i\left[\hat{\mathcal{H}}_{fS},\hat{\rho}_{fS}\right]\\
\partial_t\hat{\rho}_{bS}&=
-i\left[\hat{\mathcal{H}}_{bS},\hat{\rho}_{bS}\right] \label{eq:33}
\end{align}
We begin with the dynamics of the free species. Exchange facilitates association of free species to the catalyst ($ \hat{K}_{a,fS}\hat{\rho}_{fS}$), acting on $ \hat{\rho}_{fS} $ to remove free species from the manifold as they bind the complex, and allows for dissociation of bound ligand $ \hat{K}_{d,fS}\hat{\rho}_{bS}$, adding species back to the manifold. Both of these exchange processes happen at the exchange rate of the ligand, $ k_N $, with an action scaled proportional to the ratio of the concentration of the iridium complex to the free ligand ([Ir]/[S]) to account for the inherent trace normalization of density matrices. The association operator is then simply:
\begin{align}
\hat{K}_{a,bS}\hat{\rho}_{fS}=-\frac{\left[Ir\right]}{\left[S\right]}\hat{\rho}_{fS} \label{eq:34}
\end{align}
The dissociation operator then deposits an equivalent number of ligands from the bound species subsystem into the free species subsystem. For the case where both available binding sites in the iridium complex are exchanging with the target ligand, there are distinct subsets of the nuclear spins in the bound species ($ S_a $ and $ S_b $) which may dissociate to join the free species with equal probability. We average these possibilities to generate the dissociation operator for the free species, remembering to apply the concentration scaling factor for exchange between manifolds:
\begin{align}
\hat{K}_{d,fS}\hat{\rho}_{bS}=\frac{\left[Ir\right]}{2\left[S\right]}
\left(Tr_{\left\lbrace H_2S_a\right\rbrace}\hat{\rho}_{bS}+Tr_{\left\lbrace H_2S_b\right\rbrace}\hat{\rho}_{bS}\right) \label{eq:35}
\end{align}
Combining \eqref{eq:34} and \eqref{eq:35} yields the equation of motion for the free species with exchange interactions
\begin{align}
\nonumber \partial_t\hat{\rho}_{fS}=&-i\left[\hat{\mathcal{H}}_{fS},\hat{\rho}_{fS}\right]\\
&+\tilde{k}_N\frac{\left[Ir\right]}{\left[S\right]}
\left(\hat{K}_{d,fS}\hat{\rho}_{bS}-\hat{\rho}_{fS}\right)\!, \label{eq:36}
\end{align}
where $ \tilde{k}_N \equiv k_N\exp\left(-\Delta t k_N/2\right) $ and will be used as a notation for the DMEx rate going forward.\\
\indent The bound species has two exchange interactions, one for the simultaneous exchange of a ligand and the hydrides occurring at rate of  $ k_H $  and one for the exchange of target ligands at a rate of  $ k_N-k_H $. We will formulate the exchange operator for the bound species as a single entity,  $ \hat{K}_{ex} $, which takes multiple manifolds as arguments. Hydride exchange is restricted to occur only during ligand exchanges as the complex form a tetra-hydride intermediate to facilitate this reaction. In the case where both parahydrogen and ligand exchange occur concurrently, we exchange the portion $  \Delta t(\tilde{k}_{a,H}/\tilde{k}_{N}) $ of $ \hat{\rho}_{bS} $ to reflect the new hydride population and new ligand population. This may be written as:
\begin{align}
\left(\frac{\tilde{k}_{a,H}}{\tilde{k}_{N}}\right)\hat{\rho}_{pH_2}\otimes\hat{\rho}_{fS}\otimes
Tr_{\lbrace H_2,S\rbrace}\hat{\rho}_{bS}
\end{align}
Where $ \hat{\rho}_{pH_2} $ is the density matrix for pure singlet parahydrogen and $ Tr_{\lbrace H_2,S\rbrace} $ returns the density matrix for the ligand which remains bound. In the case where the hydrides do not exchange, but the target ligand does, another portion of the density matrix $  \Delta t((1-\tilde{k}_{a,H})/\tilde{k}_{N}) $ must be reformulated to reflect the newly exchanged ligand:
\begin{align}
\left(1-\frac{\tilde{k}_{a,H}}{\tilde{k}_{N}}\right)
Tr_{\lbrace S\rbrace}\hat{\rho}_{bS}\otimes \hat{\rho}_{fS}
\end{align}
Where $ Tr_{\lbrace S\rbrace}\hat{\rho}_{bS} $ is the density matrix for the subsystem of the remaining ligand and parahydrogen. This projection must be constructed carefully to ensure that the coherences are appropriately retained between the hydrides and remaining ligand. It is important to note that while we are exchanging between the free and bound species subsystems, the scaling factor is not needed as the free species density matrix is, by definition, trace normalized. Therefore, one free ligand equivalent leaving the free species will look like one free ligand equivalent associating with the bound species. As this free ligand leaves the free species though, the appropriate reduction in the free species density matrix must be scaled by the concentration ratios. The full exchange operators can now be written as a combination of these two components:
\begin{align}
\nonumber \hat{K}_{\lbrace S\rbrace}&\!\left(\hat{\rho}_{bS},\hat{\rho}_{fS}\right)=
\left(\!1-\frac{\tilde{k}_{a,H}}{\tilde{k}_{N}}\!\right)Tr_{\lbrace S\rbrace}\hat{\rho}_{bS}\otimes \hat{\rho}_{fS}\\
&+\left(\frac{\tilde{k}_{a,H}}{\tilde{k}_{N}}\right)\hat{\rho}_{pH_2}\otimes\hat{\rho}_{fS}\otimes
Tr_{\lbrace H_2,S\rbrace}\hat{\rho}_{bS}
\end{align}
The two possible ligand exchanges from the two available binding sites, \textit{a} and \textit{b}, then average together to give the final exchange operator for the bound species:
\begin{align}
\hat{K}_{ex}=\frac{1}{2}\left(\hat{K}_{\lbrace S_a\rbrace}+\hat{K}_{\lbrace S_b\rbrace}\right) \label{eq:Kex}
\end{align}
\indent It is critical to note that equations \eqref{eq:35} and \eqref{eq:Kex} contains terms that are quadratic in the magnetization density, arising from the effects of rebinding ligands that have already interacted with the  species. As such, this is a second order nonlinear partial differential equation, which must be solved simultaneously with the equation of motion for the free species to define the full evolution of the system. 
Furthermore, these nonlinearities are amplified as the ratio  increases. It now becomes possible to efficiently represent the impact of concurrent evolution of the \textit{J}-coupling networks in the free and bound species of the target ligand. Additionally, we can now model the effects that various solution compositions will have on the polarization dynamics, given that rebinding of previously polarized ligand will significantly impact the evolution of the bound species under the nuclear spin Hamiltonian.\\
\indent Even with the incorporation of the nonlinear terms to the DMEx, the solution convergence is still far faster than that of the traditional implementation (Fig. \ref{fig:error SABRE}C), and the two models still converge in the limit when $ \Delta t \rightarrow dt $. One can obtain the same error in the DMExFR2 with a step size that is four times larger than the traditional implementation. While we have focused on the accuracy of the simulation, its precision in reproducing input parameters, such as the exchange rate, are just as important, particularly as these models are used to extract physical parameters from experimental data. Under this condition, it is critical that the simulation is stable but also efficient, as large portions of phase space have to be searched to perform an experimental fit. To characterize the precision of the simulation, we introduce the parameter $ \sigma_k $, which defines the relative shift in the predicted exchange rate in the simulation (Fig. \ref{fig:error SABRE}D). Surprisingly, there was essentially no exchange rate at which the traditional implementation provided a solution that was stably precise. In contrast to that, the DMEx model essentially perfectly reproduces the input exchange rate until $ k \approx 300 s^{-1} $, and when nonlinearities are introduced to the simulation, the maximum deviation from the input exchange rate is only $ \sigma_k \approx  0.5\%$.\\

\begin{figure*}[!htb]
	\centering
	\includegraphics[width=\linewidth]{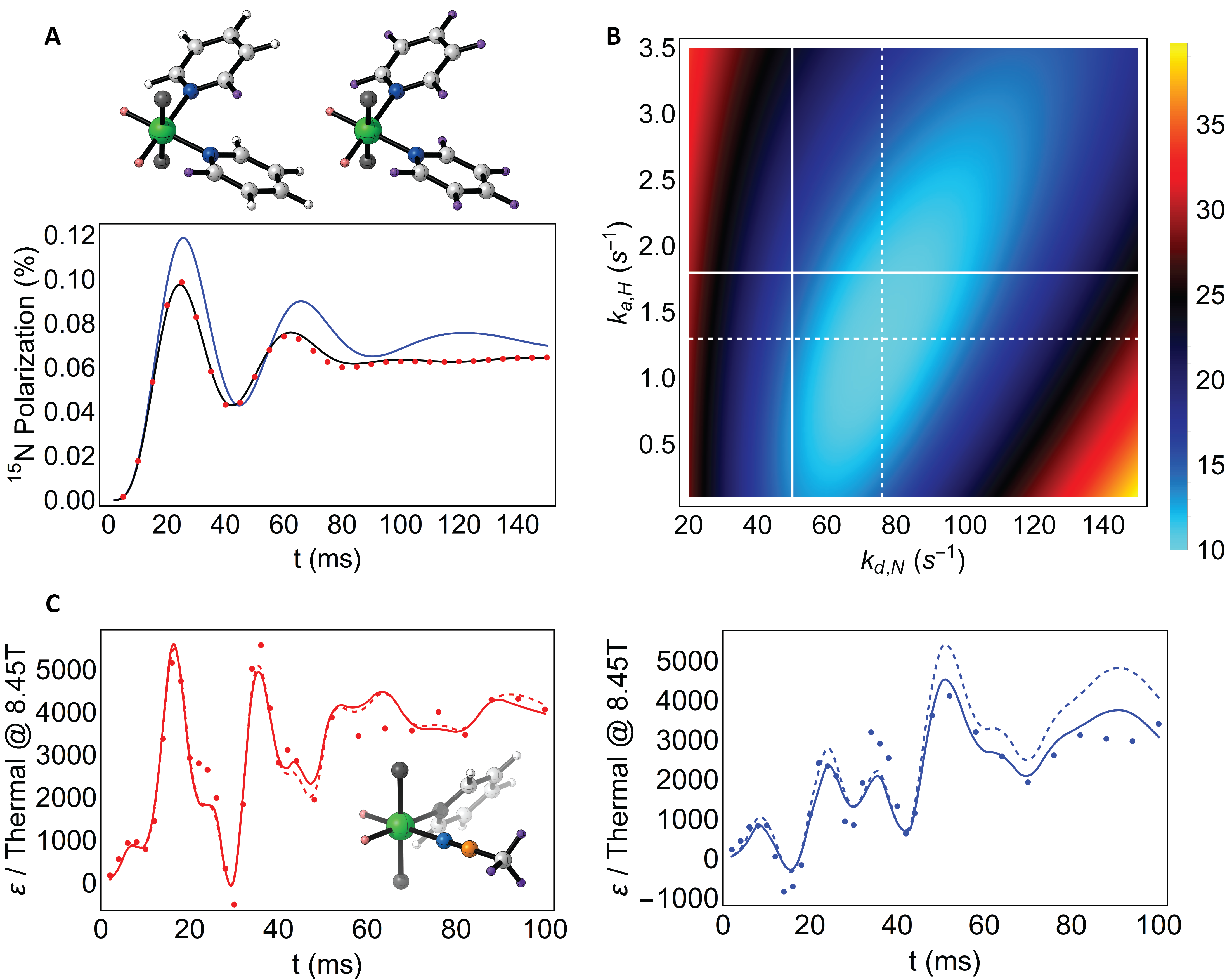}
	\caption{\textsf{Illustrative simulations of complex SABRE systems. (\textbf{A}) Truncation of the model drastically alters the hyperpolarization dynamics, as shown on a bis-($ ^{15} $N-pyridine) system with a single ancillary proton per ligand (blue) and compared to the explicit system (black) using $ \Delta t = 2 $ ms or 5 ms with only minor difference in the solution. (\textbf{B}) Truncation of the model results in different predicted exchange rates (dashed) as opposed to the actual values (solid). (\textbf{C}) Experimental data probing coherent hyperpolarization dynamics at may be fit with the DMExFR2 simulations to extract exchange rates, which can be done accounting for coligand effects (solid) or without these effects (dashed). The resonant field was $ B = -1.65$ $\mu T $ (red) or $ B = -0.91$ $\mu T $ (blue), and the predicted exchange rates using coligand effects were found to be $ k_N = (14.5 \pm 1.8) $ $s^{-1}$ and $ k_H = (6.00 \pm 0.75) $ $s^{-1}$ for $ B = -1.65$ $\mu T $ and $ k_N = (15.0 \pm 3.3) $ $s^{-1}$ and $ k_H = (4.50 \pm 0.98) $ $s^{-1}$ for $ B = -0.91$ $\mu T $. Without coligand effects, the predicted rates are $ k_N = (25.5 \pm 3.0) $ $s^{-1}$ and $ k_H = (11.5 \pm 1.4) $ $s^{-1}$ for $ B = -1.65$ $\mu T $ and $ k_N = (23.0 \pm 5.7) $ $s^{-1}$ and $ k_H = (6.5 \pm 1.6) $ $s^{-1}$ for $ B = -0.91$ $\mu T $. Relaxation is included in the model as a boundary value problem with $ T_{1,N}=20 $ s and $ T_{1,H} = 2$s. }} \label{fig:fancy}
\end{figure*}

\noindent{\textbf{\textit{Practical examples}}}\\
As noted previously, guided \textit{in silico} exploration of novel experimental methods that increase the hyperpolarization of SABRE are the focus of optimization efforts in the community. With the improved stability of the DMEx models, it is possible to explore realistic systems with complex coupling networks and reduce the calculation to an obtainable cost by utilizing large simulation step sizes ($ \Delta t > 1 $ ms). The flexibility of this formulation to be expressed in either Hilbert or Liouville space additionally provides access to much larger spin systems than previously possible.\\
\indent The case of the canonical bis-($ ^{15} $N-pyridine) SABRE-SHEATH system is particularly interesting as it contains 14 strongly coupled spins in just the ``iridium-bound'' manifold with 22 total spins and is perhaps the most prevalent system in $ ^{15} $N SABRE. As the full system is far outside the scope of previous exchange models for SABRE, it has been traditionally acceptable to truncate the spin system to an approximate system, fully or partially removing ancillary $ ^1 $H nuclei, with the largest approximation reported in literature using a single $ ^1 $H per ligand \cite{RN2138}. Even in this case, the dynamics of the truncated model diverge greatly from the actual system dynamics (Figure \ref{fig:fancy}A), the latter which can be explicitly calculated using the DMExFR2 model with either 2 ms (black) or 5 ms (red) step sizes with only minor deviation between the solutions. Resultingly, the truncated model optimizes to exchange rates that are false while retaining a deviation of $ \approx 10\% $ from the actual system dynamics when re-optimized to the erroneous rates (Figure \ref{fig:fancy}B). This means that any physical parameters extracted from experimental data by the model will be greatly confounded by the truncation errors inherent to the formulation.
\\
\indent To emphasize the efficiency and flexibility of this framework, we used the DMExFR2 model to fit the coherent hyperpolarization dynamics of ($ ^{15} $N-$ ^{13} $C)-acetonitrile when exciting the sample with short (ms) pulses tuned to a field near the SABRE resonance condition, as described in our prior work \cite{RN2474,RN2109}. Coherent evolution is then interrogated by varying the resonant pulse length, which encodes the dynamics in the final polarization detected. This is a multicomponent SABRE system containing 21 total spins and requires consideration of hyperpolarization-inactive coligand effects to accurately describe the dynamics. These effects allow for additional exchange pathways to influence the dynamics of the system. One of the most critical ramifications arising in allowing the hyperpolarizable ligand to exchange between positions on the complex and thus with which parahydrogen-derived hydride the ligand is coupled. In the limit of fast exchange, this makes the hydrides appear equivalent and would prevent the singlet order from being converted into observable magnetization. When coligand effects are included, (solid lines) the experimental data can be reproduced with high fidelity to experiment at multiple field conditions (Figure \ref{fig:fancy}C), such as when the resonant pulse is $ B = -1.65$ $\mu T $ (red) or $ B = -0.91$ $\mu T $ (blue). Furthermore, the extracted exchange rates for these data sets are $ k_N = (14.5 \pm 1.8) $ $s^{-1}$ and $ k_H = (6.00 \pm 0.75) $ $s^{-1}$ for the $ B = -1.65$ $\mu T $ data and $ k_N = (15.0 \pm 3.3) $ $s^{-1}$ and $ k_H = (4.50 \pm 0.98) $ $s^{-1}$ for the $ B = -0.91$ $\mu T $ data. When coligand effects are neglected, the predicted exchange rates can range to having errors of 44-92$ \% $.\\
\indent Properly simulating this system requires two 7 spin manifolds for the two conformers of the iridium complex, a 5 spin manifold for the free ($ ^{15} $N-$ ^{13} $C)-acetonitrile, and a 2 spin manifold for parahydrogen. Fitting the experimental data would be intractable within any of the previous formalisms for SABRE dynamics \cite{RN1648,RN2119} as a function of the system size. However, when built using the DMExFR2 model in conjunction with the manifold-diagonal formalism for exchange introduced here, each simulation data set, which consists of 32 simulations lasting 30s using $ \Delta t = 2 $ ms, requires approximately 15 minutes to calculate, making a grid-optimization fit possible within a day.\\

\noindent{\large{\textbf{Conclusions and Outlook}}}\\ 
The foundations of exchange in dynamic quantum systems has been re-assessed and derived in its exact form from infinite-order perturbation theory. In doing so, the exact Dissipative Master Equation (DMEx) formalism presented here accounts for higher moments in the ensemble action that are omitted from the traditional implementation. The speed and accuracy with which complex exchanging spin systems may be modeled using the DMEx formalism allows for extensive \textit{in silico} experimentation and optimization in a way which has previously been inaccessible. In tandem, we have introduced a ``manifold-diagonal'' implementation for exchange, allowing the simulations of multicomponent systems with nonlinear exchange dynamics to scale linearly with the dimension of the composite space. While the results described here are independent of the vector space, we have found that expressing individual manifolds in their Hilbert-space representation affords efficient simulation of experimental data with high fidelity.\\
\indent We anticipate that the results demonstrated here will have a radical impact on the simulation of complex dynamic systems. In the hyperpolarization community, the ability to accurately and efficiently simulate the entire SABRE system should greatly alter the optimization of the hyperpolarization efficiency. Annealing the DMEx formalism to state-space reduction techniques has the potential to introduce efficient simulation of systems as large as biomolecules, offering the possibility to dramatically reduce computational time and improve simulation accuracy using larger time steps.\\

\noindent{\large{\textbf{Acknowledgments}}}\\
The presented research was funded by the National Science Foundation grant CHE-1665090.

\bibliographystyle{ieeetr}
\bibliography{References}

\end{multicols}
\end{document}